# The largest structure of the Universe, defined by Gamma-Ray Bursts

I. Horvath

*National University of Public Service, Budapest, Hungary*

J. Hakkila

*College of Charleston, Charleston, SC, USA*

Z. Bagoly

*Eötvös University, Budapest, Hungary*

Research over the past three decades has revolutionized the field of cosmology while supporting the standard cosmological model. However, the cosmological principle of Universal homogeneity and isotropy has always been in question, since structures as large as the survey size have always been found as the survey size has increased. Until now, the largest known structure in our Universe is the Sloan Great Wall (SGW), which is more than 400 Mpc long and located approximately one billion light-years away. Here we report the discovery of a structure at least six times larger than the Sloan Great Wall that is suggested by the distribution of gamma-ray bursts (GRBs). Gamma-ray bursts are the most energetic explosions in the Universe. They are associated with the stellar endpoints of massive stars and are found in and near distant galaxies. Therefore, they are very good indicators of the dense part of the Universe containing normal matter. As of July 2012, 283 GRB redshifts have been measured. If one subdivides this GRB sample into nine radial parts and compares the sky distributions of these subsamples (each containing 31 GRBs), one can observe that the fourth subsample ($1.6 < z < 2.1$) differs significantly from the others in that many of the GRBs are concentrated in the same angular area of the sky. Using the two-dimensional Kolmogorov-Smirnov test, the significance of this observation is found to be less than 0.05 per cent. Fourteen out of the 31 Gamma-Ray Bursts in this redshift band are concentrated in approximately 1/8 of the sky. The binomial probability to find such a deviation is $p < 6 \cdot 10^{-6}$. This huge structure lies ten times farther away than the Sloan Great Wall, at a distance of approximately ten billion light-years. The size of the structure defined by these GRBs is about 2000-3000 Mpc, or more than six times the size of the largest known object (SGW) in the Universe.

## 1. INTRODUCTION

The cosmological origin of gamma-ray bursts (hereafter GRBs) is well established. Assuming that the Universe exhibits large-scale isotropy, the same isotropy is also expected for GRBs. The large-scale angular isotropy of the sky distribution of GRBs has been well studied in the last few decades. Most of these studies have demonstrated that the sky distribution of GRBs is isotropic (e.g. Briggs et al. 1996, Tegmark et al. 1996, Balazs et al. 1998). Furthermore, the sky distribution of the long class of GRBs is generally agreed to be isotropic (e.g. Balazs et al. 1998, Balazs et al. 1999, Meszaros et al. 2000. Magliocchetti et al. 2003).

However, some GRB subsamples appear to deviate from isotropy. Balazs et al. (1998) reported that the angular distributions of short and long GRBs are different. Cline et al. (1999) found that the angular distributions of very short GRBs are anisotropic, and Magliocchetti et al. (2003) reported that short GRB class in general deviates from angular isotropy. Meszaros et al. (2000) and Litvin et al. (2001) wrote that the angular distribution of intermediate duration GRBs is not isotropic.

In this work we examine not only the angular distribution of GRBs, but we combine this information with the burst radial distribution. As of July 2012, the redshifts of 283 GRBs have been determined (although a few of these have moderately large uncertainties[1]). This GRB sample spans a huge volume, which can presumably provide valuable information about Universal large-scale structure. In order to learn more about the properties of the Universe, we examine the Copernican principle (homogeneity, isotropy) for this sample.

## 2. THE GRB SKY DISTRIBUTION

By studying the angular distribution of GRBs as a function of distance, one can determine sample homogeneity as well as isotropy. This sample of GRBs can be subdivided by redshift, resulting in distance groupings for which angular information can also be obtained. Although the 283 GRB sample size limits our ability to set high-angular resolution limits, it can be used for lower-resolution studies. We subdivide the sample into five, six, seven, eight and nine different radial subgroups having sufficient size to justify a statistical study. Since we are examining the sample subdivided into eight and nine parts, these trivially contain the information about the sample subdivided into two, three, and four parts. Initially, we examine the sample subdivided into four radial parts (each part contain the same number of GRBs). These represent the nearest GRBs, the second nearest GRBs, the second furthest GRBs, and the furthest GRBs.

As we have already mentioned, the sky exposure function is unknown for these GRBs. Therefore it is difficult to test whether all bins have been sampled similarly. However, if one assumes similar sampling to first order, then one can test whether the two distributions are different or not. One common test for comparing two distributions is the Kolmogorov-Smirnov (KS) test. However, this test is designed to work with one-dimensional data sets; it is hard to use with data having more than one dimension, since there is no trivial way to rank in higher dimensions.

---
[1] http://lyra.berkeley.edu/grbox/grbox.php

A very good summary about how to deal with this problem is given by Lopes et al. (2008). For two-dimensional data, Peacock (1983) suggests one should use all four possible orderings to calculate the difference between the two distributions. Since the sky distribution of any objects is two-dimensional, we choose to use this method.[2]

Subdividing the sample by z produces GRB groups whose members are at similar distances from us. Or in other words, their photons come from similar Universal ages. This is not true if any group originates from a wide range in z. Therefore, the dispersion in z needs to be small. However, our sample only contains 283 GRBs. Therefore, the best way to minimize z-dispersion is to subdivide the data into a larger number of radial bins. For that reason we subdivide this sample into 5, 6, 7, 8 and 9 parts.

When we compare the five groups there is a weak suggestion of anisotropy in one group. When we compare the six groups, there is no sign for any differences between the sky distributions of the groups. That is also the case for 7 and 8 groups. But this is not the case with 9 groups.

Therefore, we focus this analysis on the nine bins containing GRBs at different redshifts. Each group contains 31 GRBs. In our case this corresponds to 31 points in the sky. The separations in z are as follows: 3.6, 2.73, 2.1, 1.6, 1.25, 0.93, 0.72 and 0.41.

## 3. TWO DIMENSIONAL KOLMOGOROV-SMIRNOV TESTS

Table I: K-S distances between the nine groups.

| no. | gr1 | gr2 | gr3 | gr4 | gr5 | gr6 | gr7 | gr8 | gr9 |
|-----|-----|-----|-----|-----|-----|-----|-----|-----|-----|
| gr1 |     | 9   | 9   | **15** | 11  | 13  | 9   | 12  | 8   |
| gr2 |     |     | 10  | **18** | 7   | **15** | 11  | 9   | 12  |
| gr3 |     |     |     | **14** | 9   | 11  | **14** | 9   | 10  |
| gr4 |     |     |     |     | **15** | 10  | **15** | **17** | 11  |
| gr5 |     |     |     |     |     | 13  | 13  | 8   | 10  |
| gr6 |     |     |     |     |     |     | 10  | 13  | 8   |
| gr7 |     |     |     |     |     |     |     | 10  | 10  |
| gr8 |     |     |     |     |     |     |     |     | 11  |
| gr9 |     |     |     |     |     |     |     |     |     |

---

[2] Note there are two other problems. 1, The plane and the sphere have different topologies. 2, One can use other coordinate systems.

Using the Peacock methodology to compare two distributions, we calculate the numbers in each quadrant in each group. When comparing two groups with 31 members, that means that there are 62x62=3844 division points, therefore one has 4x3844 numbers in each group. For these 15376 pairs, one has to find the largest of their differences. Comparing the two farthest groups the largest numerical difference is 9. Comparing the two nearest groups, the largest numerical difference is 11. For the moment, we do not know exactly what these numbers mean. However, we can compare them with one another. Table I contains the largest number in the quadrants for each comparison. Larger numbers indicate larger differences between the two groups being compared. Of the six largest numbers, five belong to Group 4. Out of the eight largest numbers, six belong to Group 4. In other words, six of the eight numbers (out of 36) measuring the largest differences between group pairs belong to Group 4.

One can calculate approximate probabilities for the different numbers using simulations. We run 40 thousand simulations when 31 random points are compared with 31 other random points. The result contains the number 18 twenty-eight times and numbers larger than 18 ten times. Therefore, the probability of having numbers larger than 17 is 0.095%. The probability of having numbers larger than 16 is p= 0.0029, of having numbers larger than 15 is p= 0.0094, and of having numbers larger than 14 is p=0.0246. For a random distribution, numbers larger than 14 correspond to 2 sigma deviations and numbers larger than 16 correspond to 3 sigma deviations. The probability of having numbers larger than 13 is p=0.057, or 5.7%, which we do not find to be statistically significant. Comparisons of the 9 groups to each other using the 2D K-S test are shown in Table I.

We have two three-sigma angular anisotropy signatures. In both cases group 4 is involved. We also have eight two-sigma signatures. In six cases group 4 is involved. However, we do not have a three-sigma signature since we had 36 different pairs to compare. Among 36 different tests one expects 1.64 two-sigma signatures and no (0.09) three-sigma signatures. Except for cases involving Group 4 we find numbers similar to these random distributions (2 two-sigma signatures and no three-sigma signatures). However the 36 comparisons are not independent since we have only nine groups to compare.

## 4. NEAREST-NEIGHBOR STATISTICS

One can also look for anisotropies using nearest-neighbor statistics. Assuming again that the sky exposure is independent of *z*, one can compare the distributions with one another. Since we are not focusing on pair correlations we should calculate not just the nearest-neighbor distances, but also the second, third etc. nearest neighbor distances. For all nine groups we calculated the k-th (k=1, 2, … 30) nearest neighbor distance distributions. Since these are one-dimensional distributions, a simple Kolmogorov-Smirnov test can be used for test. For eight groups we do not find significant deviation. But a KS test for Group 4 shows significant deviations from angular isotropy starting with the sixth-nearest neighbor pairs (figure 1. shows the 12th-neighbor distribution).

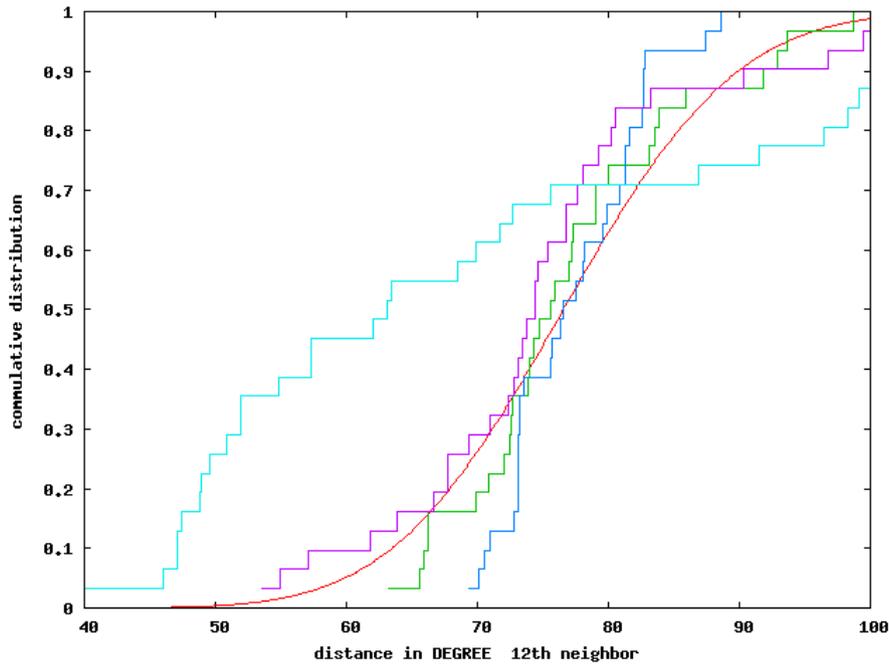

Figure 1: The 12th-neighbor distribution for the first four groups, light blue is group4.

The KS probabilities can be seen in Figure 2. For comparison we also plotted the Group 5 probabilities. One can see 21 consecutive probabilities in Group 4 reach the 2σ limit and 9 consecutive comparisons reach the 3σ limit. Of course this does not mean a 27-sigma limit, because the comparisons are not independent. One can calculate bimodal probabilities. For example 14 out of the 31 Gamma-Ray Bursts in this redshift band are concentrated in approximately 1/8 of the sky (Figure 3). The binomial probability to find such a deviation is p=0.0000055.

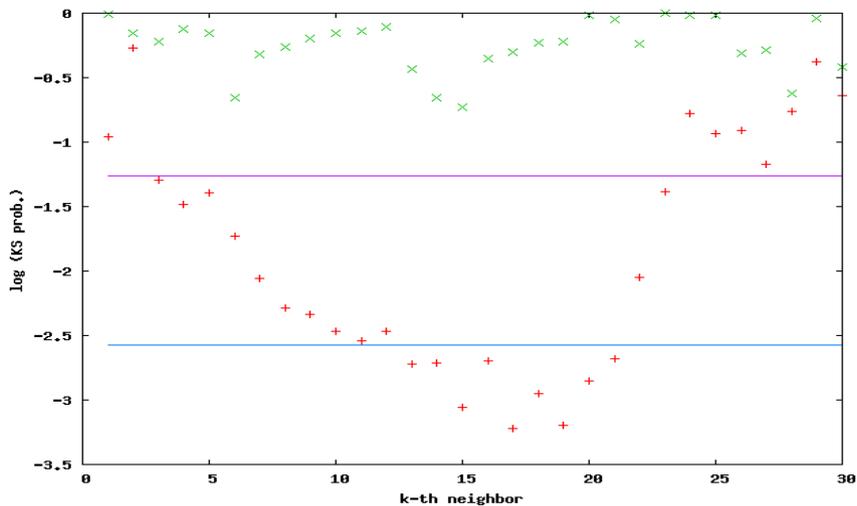

Figure 2: K-S neighbor probabilities for group4 (red) and group5 (green). Pink (blue) line indicates the two (three) sigma level.

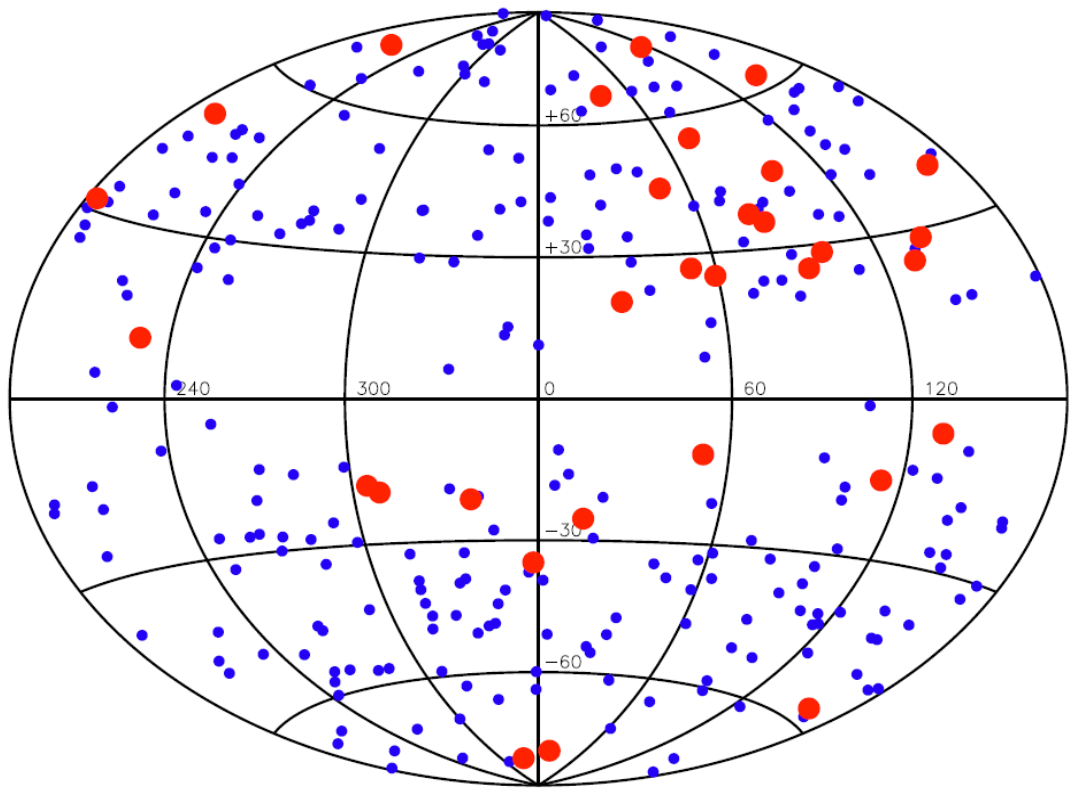

Figure 3: 283 GRBs with observed redshift (blue) and the 31 GRBs (red) between redshift 1.6 and 2.1.

## 5. ESTIMATED EFFECTS OF SKY EXPOSURE

Before taking the aforementioned angular anisotropies at face value, one should consider the possibility that an angular sampling bias (e.g. *sky exposure*) might have contributed to them. The primary cause of sky exposure is a detection bias: GRB detectors often favor triggering on GRBs in some angular directions over others.

An anisotropic sky exposure can result when a pointed spacecraft spends more time observing some sky directions than others. It can also result from blockage of the angular field-of-view, say from Earth occultation or avoidance of the Sun to protect sensitive instrumentation, or from spacecraft insensitivity over certain Earth locations (such as the South Atlantic Anomaly). Each GRB instrument experiences different degrees of sky exposure, which makes the summed sky exposure difficult to identify for our heterogeneous GRB sample observed by many instruments since the late 1990s. However, 214 of the 283 GRBs in our sample (75.6%) have been observed by Swift, as have 23 of the 31 GRBs in Group 4 (74.2%). Thus, we assume that Swift's sky exposure is a reasonable first-order approximation to the sky exposure of the entire burst sample. Swift's sky exposure has recently been published (Baumgartner et al. 2012) and is shown in Figure 4.

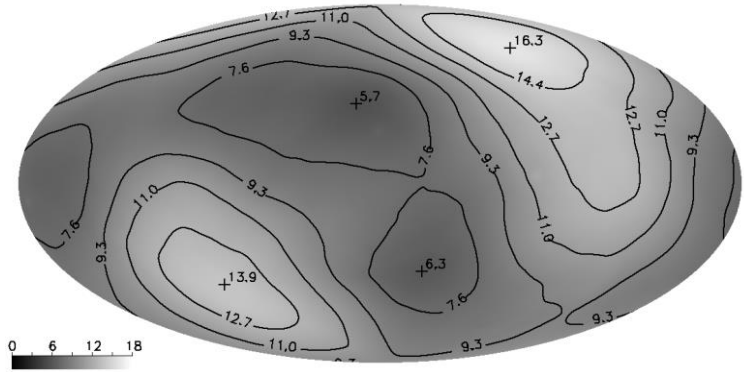

Figure 4: Swift Sky Exposure (From Baumgartner et al. 2012). High exposure regions are light-colored.

We test the reasonableness that Swift's sky exposure might have contributed to the burst excess in Group 4 by examining bulk exposure properties. Swift's sky exposure is primarily a function of ecliptic latitude, with greater exposure in the directions of the ecliptic poles relative to the ecliptic equator. In order to minimize effects of sample size in our analysis, we subdivide Group 4 into two angular regions: the polar region (NEP = north ecliptic pole; ecliptic lat. > 45° plus the SEP = south ecliptic pole; ecliptic lat. < -45°) and a mid-ecliptic latitude region (near the ecliptic equator (EE; between -45° and 45°). The 40% of the sky near the ecliptic poles has been sampled 2.5 times more frequently than the 60% of the sky near the ecliptic equator. Thus, only six bursts should have been observed near the ecliptic equator in the time that ten triggered near the ecliptic poles. Instead, 21 were found (Table II). This corresponds to an excess significant at the 7.5σ level, which is quite unlikely. To first order, sky exposure cannot explain why there are so many GRBs at z=2 near the ecliptic equator.

Table II: The sky exposure cannot explain the find of GRB grouping.

| Region | Observed | Fractional area | Exposure (relative to ecl. eq.) | Expectation (based on ecl. poles) | Uncertainty | Pull |
|---|---|---|---|---|---|---|
| ecl. poles | 10 | 0.4 | 2.5 | 10 | ± 3 | |
| ecl. eq. | 21 | 0.6 | 1.0 | 6 | ± 2 | 7.5σ |

Other sky exposure issues still must be considered. First, the angular exposure for the 25% of bursts triggering by detectors other than Swift has not been identified. This will likely be small, and will possibly be smeared out by summing the exposure of instruments with different angular biases, but it is unfortunately difficult to characterize. Other perhaps important angular biases are present in this GRB sample that mimic sky exposure. The first of these is a bias away from the galactic plane, where dust extinction prevents optical counterparts from being identified. The second is a potential bias towards northern (or southern) hemisphere observations, based on the availability of ground-based telescopes performing follow-up observations. A better characterization of angular biases is underway.

## 6. CONCLUSION

Here we report the discovery of a possible large-scale Universal structure at a distance of approximately ten billion light-years (redshift ~2). The Cosmological Principle (the assumption that the Universe is homogeneous and isotropic) is widely accepted. However, the cosmological principle of Universal homogeneity and isotropy has always been questioned, since structures as large as the survey size have been consistently found as the survey size has increased. In the late 1980s Geller and Huchra (1989) mapped the Universe to z~0.03 and found a 200 Mpc size object, which they called the "Great Wall." In 2005 an object twice this size was reported and named the Sloan Great Wall (Gott et al. 2005). In this study we have found a potential structure, mapped by GRBs, of about 2000-3000 Mpc size. One or two years more of gamma burst observations will hopefully provide the statistics to confirm or disprove this discovery.

### Acknowledgments

This research was supported by OTKA grant K77795 and by NASA ADAP grant NNX09AD03G.